# Spatial Knowledge Acquisition in Virtual and Physical Reality: A Comparative Evaluation

Diego Monteiro
*Department of Computing*
*Xi'an Jiaotong-Liverpool University*
Suzhou, China
D.Monteiro@xjtlu.edu.cn

Xian Wang
*Department of Computing*
*Xi'an Jiaotong-Liverpool University*
Suzhou, China
Xian.Wang17@student.xjtlu.edu.cn

Hai-Ning Liang*
*Department of Computing*
*Xi'an Jiaotong-Liverpool University*
Suzhou, China
HaiNing.Liang@xjtlu.edu.cn
*Corresponding author*

Yiyu Cai
*School of Mechanical and Aerospace Engineering*
*Nanyang Technological University*
Singapore
MYYCai@ntu.edu.sg

*Abstract*—Virtual Reality (VR) head-mounted displays (HMDs) have been studied widely as tools for the most diverse kinds of training activities. One particular kind that is the basis for many real-world applications is spatial knowledge acquisition and navigation. For example, knowing well by heart escape routes can be an important factor for firefighters and soldiers. Prior research on how well knowledge acquired in virtual worlds translates to the real, physical one has had mixed results, with some suggesting spatial learning in VR is akin to using a regular 2D display. However, VR HMDs have evolved drastically in the last decade, and little is known about how spatial training skills in a simulated environment using up-to-date VR HMDs compare to training in the real world. In this paper, we investigate how people trained in a VR maze compare against those trained in a physical maze in terms of recall of the position of items inside the environment. While our results did not find significant differences in time performance for people who experienced the physical and those who trained in VR, other behavioral factors were different.

## I. INTRODUCTION

Google Earth is used to show different places to people who cannot go there by themselves. Given the popularity of online exploration used in many fields (e.g., education, travel, design, training), questions about people's ability to recall the places they visit virtually are of growing importance. For instance, can the virtual experience ever create a memory like walking down the streets physically? Could people remember a route if they just walked through it via an application like Google Earth, especially in virtual reality (VR)?

One technology that promises accessible training for the most various purposes is VR. Training is arguably one of the most important functions of VR. However, since VR is a relatively new technology, there is still much which we do not know about it in terms of its effect on memory recall, cognition, and knowledge acquisition. For example, high immersion levels might cause cognitive overload [1], while lower levels of immersion might lead people to be discouraged by the technology if they are not engaged enough. Similarly, the technology brings other inherent issues, like motion-induced sickness, that are less prevalent in the physical environment and could negatively impact people's ability to make sense, think, and learn [2]–[4].

It is important to differentiate a high level of immersion with high-level training [1]. Most current studies tend to focus on higher levels of training by exploring real-life scenarios and trying to see whether VR can be helpful for these scenarios [5], [6]. In this sense, not many studies investigate in-depth how the building blocks of learning compare in real life versus in the virtual environment [7]. Studies have shown that when navigating a virtual maze, the brain shows higher activity levels when the same maze is in 3D rather than in 2D [8]. This finding indicates that changes in visualization affect how spatial memory is processed. Since VR is a highly immersive medium, it is likely that, when using VR for navigation, information can be retained in a similar fashion as in the real world.

Several aspects of VR can be used to elicit realism, such as adding a virtual character [9], increasing the resolution of the display, or allowing people to stay longer in VR by using different viewing perspectives to make their experience less motion sick [10], [11]. However, it is difficult to deeply understand the connection between realism and training if we do not understand their basis. In this study, we aim to shed some light on how similar spatial memory acquisition in VR and real life is. To this end, we conducted an experiment with a maze which allowed us to compare both environments and observe if, for short tasks, spatial learning in VR and real-life yielded similar results.

## II. RELATED WORK

### A. Training in VR

Training is potentially one of the most relevant facets in the use of VR HMDs. Some studies explored their use to train people to speak better towards audiences [12]–[14]. Other studies looked into them in more traditional training environments, such as classroom teaching of science and biology at different levels [15], [16]. Others have investigated their effectiveness in navigation tasks, for example, in firefighting, preflight training for astronauts, and remembering shop locations [17]–[19].

Lee et al. [20] investigated how to instruct better people on memorizing paths in VR. To do that, they compared the difference between annotating the path and following a virtual character. They found that people performed better when they were instructed on what to do rather than to follow a character. This effect might have to do with some form of excessive mental demand similar to that found when advanced biology students were studying in the virtual lab [21]. It also might suggest that further developments on character are not needed for this kind of context training [9].

Earlier studies of spatial memory acquisition compared 2D and 3D mazes. They used a computer screen against a CAVE-like environment and observed that there was a greater level of certain brain activities when participants were learning the paths in 3D when compared to the participants learning it in 2D. This observation shows that the greater the realism, the better it might be for spatial learning [22]–[24].

Suma et al. [25] compared a maze viewed in VR HMD and in real life to evaluate different kinds of locomotion. They observed that people felt sicker when walking in the VR environment than when moving using a controller. However, people had better recall when walking. However, this study was done in 2009, before low-cost, higher-resolution HMDs became accessible, which could have changed this paradigm. As such, evaluations with newer HMDs are still needed to explore this issue further.

Recently, Mossel et al. [6] investigated the differences between using a controller and a walking platform in a firefighting scenario and a mobile VR device. The participants in this study did not favor one kind of locomotion over the other. However, there was an emphasis on the importance of navigation being a "*very important*" factor in these kinds of drill scenarios.

Furthermore, the interaction between view and how the virtual environment is controlled can be of significant effect on performance and immersion. In a recent study, Monteiro et al. [26] observed that in a 2D view environment, participants performed better in their task when they were using a keyboard than the regular controller; on the other hand, in 3D view, the controller fared better than the keyboard.

Finally, in a study that evaluated what aspects of locomotion were more likely to affect the learning outcomes in a maze, Chrastil and Warren [27] observed that visual and podokinetic (or foot movement) information significantly contribute to path memorization. It is likely that, in a scenario that involves real locomotion, having experienced the controller without the associated foot movement and other aspects of the controller not being as natural might lead to lower path memorization. This observation led us to our first hypothesis:

*H1 – Participants will perform better after training in the real world than when training in VR.*

Further, because of the lack of one modality of sensory input, which could lead to greater difficulties in VR, we extrapolate a second hypothesis we will investigate:

*H2 – Participants whose navigation is poor will be slower after training in VR than those after training in the real world.*

That is, we believe that even if the first hypothesis is not supported by our data, a subgroup of our participants will present the expected behavior.

*B. Video Games and Learning*

In general, studies have shown that video games can alter cognition and perception. A typical example is, for example, people who, after being exposed to video games, consistently present improvements in mental rotation [28], [29]. Other examples include people who show improvement in vision-related challenges, like better performance in search tasks, attention, and muscle memory, also show greater sensitivity to contrast [30]–[35]. To some extent, some of these aspects have also been shown in VR environments [36]–[38].

More recent studies have suggested that video games might not enhance navigational skills in the real world. This is aligned with the findings from Chrastil and Warren [27], who observed that foot movement information significantly contributes to path memorization. That is, because video games do not involve foot movement, path memorization will not be affected. Nonetheless, in the same work, Chrastil and Warren [27] have found that visual input is an important factor in path recall. As such, because of the enhanced immersive experience in VR, we hypothesize that VR will have a mediating effect, which is our third hypothesis:

*H3 – Participants who play video games regularly will score better than those who do not.*

### III. EXPERIMENTAL DESIGN

To test our hypotheses, we used a maze challenge with which to run our experiment. We chose a maze because of its flexibility, allowing us to decide the degree of difficulty we would explore. Additionally, mazes are, to a large extent, one of the most basic forms of navigation scenarios, working as placeholders for buildings and streets, for example. In other words, it is the simplest form of pathfinding and navigation with which people are familiar. Thus, a maze grants us an advanced understanding of the building blocks of spatial knowledge acquisition before exploring higher dimensions of learning.

Our maze was built using portable and modular empty frames that connect to each other to generate more complex structures. This gave us the ability to transport and remodel the maze according to need. The walls were made using a black ethylene-vinyl acetate (plastic) fabric. The dark and non-transparent fabric stopped participants from spying into other parts of the path while not being heavy (as seen in Fig 1). The maze was designed in association with a collaborator from the architecture department to guarantee that it would fit in the room and that the proportions in the VR environment would be adequate.

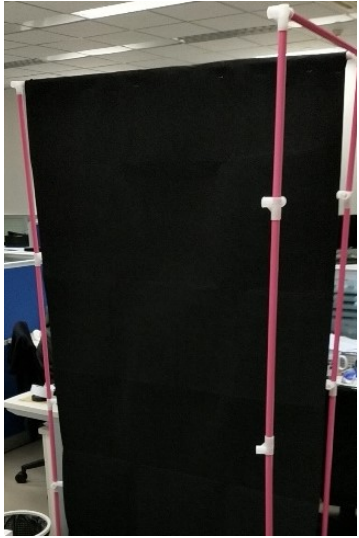

Fig 1. Picture of our maze material. The ease to assemble was vital to an indoor maze.

As presented in Fig 2, we distributed the objects in a way that would require the participants to get into distinct depths of the maze for retrieving them. We expected the participants to have at least three decision points before reaching the destination (i.e., each item). The virtual maze was developed using Unity3D, following the instruction given in the blueprint and assuring that the virtual walls' size matched the size of the physical walls in a 1-to-1 fashion.

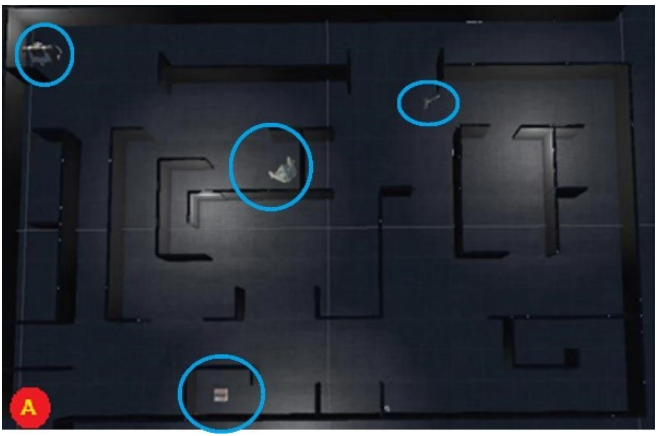

Fig 2. Picture of the distribution of objects. The letter A represents the beginning and the end of the maze. The blue circles indicate the position of the items used in this experiment.

### A. Apparatus

We used an Oculus Rift CV1 as our HMD, as it is one of the most popular off-the-shelf VR devices. The HMD was connected to a desktop with 16GB RAM, an Intel Core i7-7700k CPU @ 4.20GHz, a GeForce GTX 1080Ti dedicated GPU, and a standard 21.5" 4K monitor. We used a traditional Oculus Touch, which is one factor we are collecting data on in this experiment.

### B. Participants

We recruited a total of 10 participants (4 females) from a local university. They had an average age of 21.10 (s.e. = 3.73), ranging between 18 and 30. All volunteers had normal or normal-to-corrected vision, and none of them declared any history of color blindness or other health issues, physical or otherwise. Nine participants (90%) had experience with VR systems before the experiment. The other participant declared having both poor navigation skills and do not play any kind of video games that involve navigation in a virtual environment.

Participants were assigned randomly to each condition, respecting the rule that each condition would have the same number of participants, and male to female ratio. No participant received any kind of compensation, monetary or otherwise. All participants were made aware of the experimental conditions and expectations and consented to the experiment.

### C. Procedure

Fig 3 shows the overview of the experiment. On arrival, each participant was assigned the specific condition in which he or she would test the maze. The random assignment had been chosen to mitigate any kind of bias. The participants then filled a questionnaire to collect demographic and past gaming experience information, such as age, how often they played games, whether they had experience with VR, and their navigation skills.

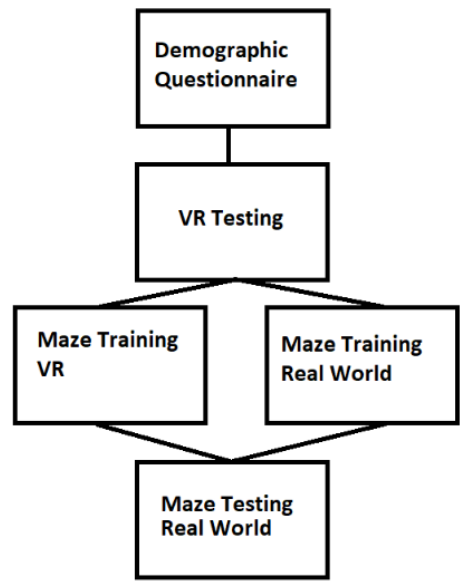

Fig 3. A flowchart of the experimental design.

Each participant stood individually next to the computer in which the virtual maze was installed. A researcher came and explained how the VR equipment data collection worked so that the participants would be more comfortable wearing the device. After the explanation, the researcher positioned the device on the participants' heads, checked the comfort level, and repositioned the device if necessary. All participants then tested the device with the Oculus Demo scene for 2 minutes so that participants in the VR condition would know how to deal with the Oculus Touch, while the participants in the real-life (RL) condition would have had the same exposure to ensure that the extra VR training would not be a confounding factor.

After the introduction, we asked participants to wait 5 minutes to give them a rest from the VR HMD. Subsequently,

they would then start the training. This time the VR HMD was already set for the participant, and the researcher had only to check that the devices were positioned correctly. Participants who did not use VR were just asked to start their walk in the maze. Both groups were timed to ensure they would not overextend their time within the training.

We required the participants to take 10-minute breaks between training and testing. Within these 10 minutes, they were asked to play a shooting game on a separate computer and solve a series of small math exercises so they would have their minds out of the maze. We wanted to ensure that they did not have a chance to keep looking at the maze, memorizing the path, giving them an unfair advantage. We made sure the waiting time was exactly 10 minutes for all participants.

During the test, one researcher would tell the participant which object should be retrieved from the maze. After this, the researcher started the stopwatch for that specific object. The participant should, at the lowest time, bring that exact object back to the researcher. The participants were not allowed to run and were not informed of the time of other participants so as not to foster competition which could skew the results. Even though researchers could not see into the maze, they could observe the first directions participants took, and those directions were written down. Each trial in the maze lasted approximately 1 minute and 20 seconds.

Overall, the experiment lasted around 35 minutes for each participant. Towards the end, they were offered refreshments such as beverages and snacks and were asked to stay in the lab for a few minutes to make sure they were not impaired or uncomfortable due to possible simulator sickness or general fatigue from exercising. Finally, they were informed that they could leave the study lab at any moment if they so desired and feel comfortable doing so.

## IV. RESULTS

The data were analyzed using both statistical inference methods and data visualizations. We conducted a Shapiro-Wilk test to check the normality of the data. As all were classified as normally distributed, they received a parametric analysis. We conducted Mauchly's Test of Sphericity to verify if the assumption of sphericity had been violated. We also employed Repeated Measures ANOVA (RM-ANOVA) using Bonferroni correction to detect significant main effects. If the assumption of sphericity was violated, we used the Greenhouse-Geisser correction to adjust the degrees of freedom.

For the analysis, we compared VR and real-life (RL) navigation in the maze. However, given a clear divide between participants who declared themselves as experienced in games, we also checked to see if there was a clear difference between those who were experienced and those who were not.

Fig 4 shows a summary of the main results. ANOVA tests suggested that there were no significant main differences in Total Time ($F(1, 8) = .082, p = .781$) between the two conditions. We further analyzed each time individually to observe possible individual differences due to the positioning of the items. However, none of the times showed any difference either (for all $p>0.05$).

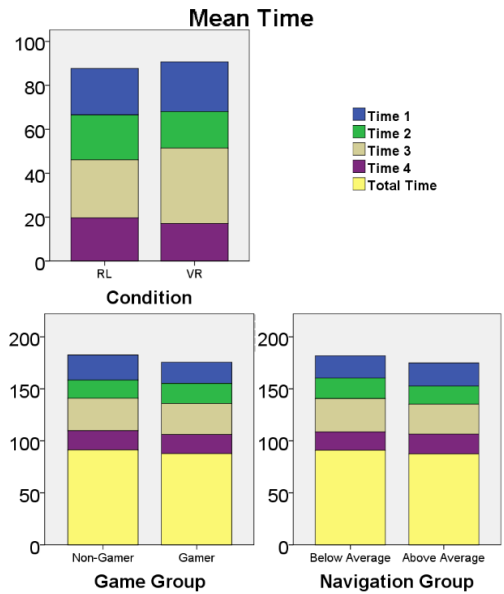

Fig 4. Mean times by condition and demographics.

The analysis taking into consideration the experience level people have from playing 3D games also did not present any significant differences in Total Time ($F(1, 8) = .140, p = .718$). To be thorough, we investigated each individual's times as well. The individual times did not present significant differences either (for all $p>0.05$). When checking if the declared level of navigation ability had an influence on time, no significant difference was found in Total Time ($F(1, 8) = .140, p = .718$). We were unable to discard the null hypothesis in *H1*, *H2*, and *H3*.

When going for the third item, most (4 out of 5) of the VR participants chose going right at first and then turning, whereas most (4 out of 5) of the participants who walked in the RL maze first went straight and then turned right (see Fig 5).

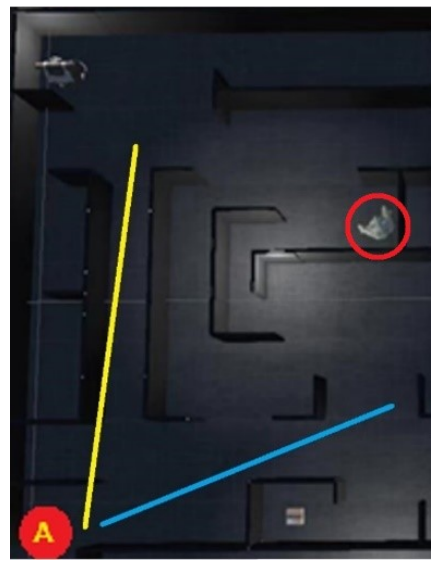

Fig 5. The yellow line represents the beginning of the route to retrieve Item 3 taken by the participants after training in the real-life condition. The blue line shows the path taken by the participants after training in VR.

## V. DISCUSSION

We believe these results to be very positive to those who want to use VR as a training tool since, regardless of expertise level with video games, the times of spatial recall using the VR were the same as the time of recall of those who used RL. This is a step up from what has been presented in previous studies that used older VR technology [25]. This older study did present positive results, but VR was still not on par with RL.

It is possible that the smaller number of items in our study allowed our participants to remember the path better. However, this demonstrates that if people are fed in small pieces, training in both VR and RL yields virtually the same results.

It is interesting to observe that the participants chose different paths so consistently when going for an item located deeper into the maze. First, this might explain the slightly longer time participants took during this item's retrieval process. Second, it might indicate that visual memory and podokinectic memory regard distance differently, emphasizing the need for studies evaluating gaze tracking in VR [39], [40]. It is possible that the participants took straighter paths inside the maze using the predicted longer route, whereas the virtual reality participants did not have this "feeling" of distance. Naturally, we are intrigued why VR participants overwhelmingly chose one path, whereas the RL participants chose the other. To understand this pattern further will require more research to gain additional insights into the difference between the two path choices. While unlikely, it can be caused by other small clues, such as imperfections on the walls, the researcher's body language. However, it is important to note that the researcher responsible for the maze did not know the participant's condition, and the material on the wall was new. This might indicate that distinct parts of the brain are involved in each process, which could mean that VR could eventually surpass RL training for spatial knowledge acquisition. Further research is required to confirm this hypothesis, given that such in-depth investigations are beyond the scope of this paper.

One limitation of this research is the size of our sample, as our experiment was affected by Covid-19. However, our number of participants is aligned with research presented in most papers published at one of the most rigorously reviewed conferences, CHI [41]. As such, it does not invalidate our scientific contribution and our findings. What we learned could be used to inform future research, which is that spatial knowledge acquisition is less likely to be the cause of confounding factors. That is, when testing navigability, path training, and other spatial cognition, we can infer that if no other variables are being tested, the results will be like that of the real world or quite similar to it.

## VI. CONCLUSION

In this paper, we explored if spatial memory acquisition was comparable between real life (RL) and virtual reality (VR). To do so, we developed a maze experiment in which participants had to get into a maze and retrieve objects from it in an orderly way. Participants performed roughly the same in VR as they did in an RL maze of the same features regardless of their navigational ability and video game playing habits.

Our results show that current VR technology can be confidently used for various applications that involve learning paths and memorizing locations, at least for tasks in short paths. It allows future research to discard negative results in spatial memory acquisition as a VR HMD cofound factor. It is important that research like this is continually explored using the latest VR technologies, as they change rapidly and in significant ways, and is published so all posterior research could have empirical support to explore more complex investigations and scenarios.

To the best of our knowledge, this is the first work that investigates the differences between RL and VR in spatial knowledge acquisition using modern consumer-grade VR HMD. Future steps involve analyzing other maze components like the point of view, bird eyes and if the use of kinematic controllers can improve the results further.

## ACKNOWLEDGMENT

The authors would like to thank the participants for their time and the reviewers for their reviews. The work is supported in part by Xi'an Jiaotong-Liverpool University (XJTLU) Key Special Fund (KSF-A-03; KSF-P-02) and XJTLU Research Development Fund.

Authors' background (This form is only for submitted manuscript for review) *This form helps us to understand your paper better, the form itself will not be published. **Please delete this form on final papers.** *Title can be chosen from: master student, PhD candidate, assistant professor, lecture, senior lecture, associate professor, full professor.

| Your Name | Title* | Affiliation | Research Field | Personal website |
|---|---|---|---|---|
| Diego Monteiro | PhD Candidate | Xi'an Jiaotong-Liverpool University | HCI, VR/AR, Games | |
| Xian Wang | Undergraduate Student | Xi'an Jiaotong-Liverpool University | HCI, VR/AR, Games | |
| Hai-Ning Liang | Senior Associate Professor | Xi'an Jiaotong-Liverpool University | HCI, VR/AR, Games | |
| Yiyu Cai | Associate Professor | Nanyang Technological University | VR, AR, Education Technology, Games | |